# Un référentiel spatial pour la gestion de la voirie dans son environnement

Antonin Pavard, Anne Dony, Patricia Bordin

Pas un jour vous ne la foulez ou ne roulez dessus. Vous y êtes tant habitués qu'elle se soustrait à votre attention : c'est … la voirie routière ! Cet oubli n'est pas propre aux usagers. L'architecte Alonzo (2018a) relevait que l'agence d'urbanisme chargée de l'aménagement de l'A10 à Tours disposait de plans sur lesquels la voirie était laissée blanche. Le même Alonzo (2018b) présentait lors d'un congrès la tendance à effacer la voirie des discours publicitaires sur les nouveaux modes de transport. Bien qu'oubliée, la voirie réapparait lors d'incidents de circulation tels que les bouchons ou les accidents (Qi 2021), ou lors de dégradations telles que les nids-de-poule (Le Figaro.fr 2018 ; Palacin 2018 ; Razemon 2018). Ainsi, le plus souvent, la voirie se rappelle à nous dans son rôle de support des mobilités et d'ouvrage participant à la sécurité des riverains ou à l'aménagement de la ville.

Pour autant, elle est bien plus complexe que cela, de par ses multiples dimensions : par exemple politique, lorsqu'il est question de gérer son patrimoine, ou technique pour la gestion mécanique de l'ouvrage. Et c'est en milieu urbain que sa complexité est plus accrue. Par exemple, une plus grande diversité de modes de transport y circule. En France, le retour d'anciens modes de transport (tramway) a d'ailleurs conduit à reconsidérer son développement axé autour de l'automobile (Héran, 2015 ; Papon et De Solère, 2009). Pour cette raison, des acteurs de la voirie tels que le législateur, les aménageurs et urbanistes tentent de rendre interopérables les modes de transport en repensant l'infrastructure support (Héran, 2017 ; Augello, 2001).

La voirie accueille également une diversité d'activités telles que les marchés, la végétation pour rafraichir ou embellir la ville, ou encore des réseaux techniques (eau, gaz ou électricité). L'inscription de ces activités complexifie la voirie par l'ajout d'interventions d'acteurs de domaines variés (de la route, des réseaux techniques, ou des commerces). Cette diversité d'acteurs nécessite de bien connaitre la voirie et ce pour sa gestion (aménagements, coûts, etc.) et notamment sa maintenance (Faivre d'Arcier et al., 1993). En 2017, les assises nationales de la mobilité ont indiqué que les gestionnaires doivent mieux connaitre leurs voiries, et améliorer la communication entre les acteurs concernés (MTES, 2017). Ceci n'est possible, selon nous, que par la conception d'un référentiel commun permettant de représenter et décrire la voirie dans son environnement. Il convient pour cela de s'interroger sur plusieurs points : qu'est-ce que réellement la voirie ? comment la représenter spatialement ? comment la décrire techniquement ? Nous avons traité ces problématiques à travers l'exemple de la cogestion de voirie et des réseaux techniques enterrés (Pavard 2020).

**Qu'est-ce que la voirie routière ?**

Les acteurs impliqués par des enjeux variés impactent la voirie à différents niveaux. Par exemple, les choix de construction dépendent des activités qui s'y déroulent : pour inviter le public à participer aux activités d'un secteur de la ville, les aménageurs en interaction avec les gestionnaires et les entreprises routières proposent un découpage de la voirie pour des usages ciblés (espaces réservés aux véhicules, espaces piétonniers, etc.). Certaines activités ont aussi un impact sur la qualité de l'infrastructure : l'enfouissement des réseaux techniques entraine des interventions de maintenance par excavation de la voirie, qui réduisent la durée de vie estimée de l'infrastructure. Enfin, la voirie est aussi marquée par la diversité d'échelons administratifs (état, départements, intercommunalités, communes) qui en ont la gestion, ainsi que du milieu dans lequel elle est implantée (urbain, péri-urbain, ou rural).

Les différents besoins et les sensibilités professionnelles de ces acteurs sont à l'origine d'une hétérogénéité des visions de la voirie. Lors de nos échanges avec ces acteurs (gestionnaires de voirie et de réseaux), et au travers de nos lectures, nous avons observé une grande richesse et variabilité du vocabulaire utilisé pour décrire la voirie. Chacun peut la désigner par un terme différent, entrainant des confusions. Nous avons relevé des exemples lors de réunions ou de présentations en séminaires professionnels :

– L'usage de deux termes pour une même emprise, tels que « route » et « chaussée » pour désigner la partie utilisée pour la circulation des véhicules (**Figure 1 – a)** ;
– L'usage d'un même terme pour deux emprises différentes, tel que « voirie » qui renvoie soit à l'emprise utilisée par l'automobile (CERTU 2013), soit à la même emprise augmentée des trottoirs (Merlin 1985) (**Figure 1 – b**).

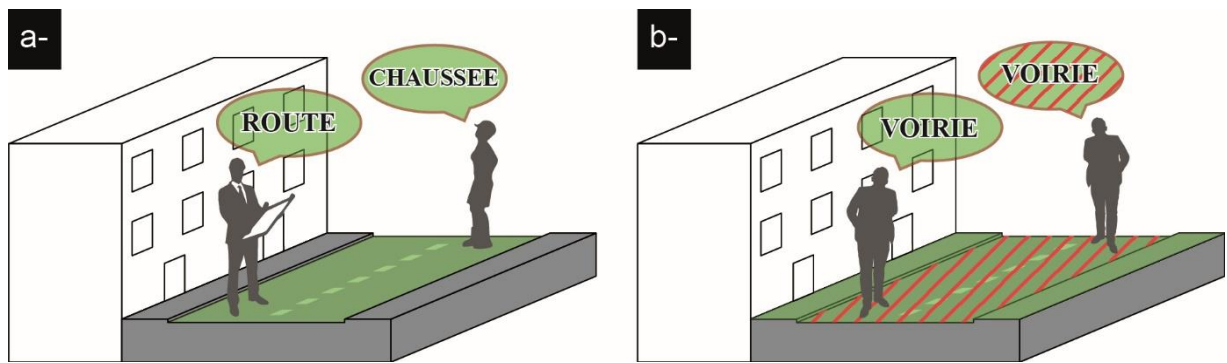

*Auteur : Pavard 2021*

**Figure 1 – Terminologie pour désigner la voirie**

Cette diversité d'usage des termes s'explique par l'histoire de la voirie et les spécialisations des activités supportées qui se sont opérées en même temps que sa complexification. **De ces constats, une question émerge : comment construire une vision harmonisée de la voirie ?** Selon nous, cela est d'abord réalisable par l'uniformisation des termes. Nous avons alors travaillé sur la construction d'un lexique commun à partir d'une analyse sémantique de textes historiques, règlementaires et techniques traitant de la voirie. La **Figure 2** présente un extrait de notre lexique qui repose sur la définition de l'objet au cœur du sujet, la voirie, et que nous définissons comme l'« *infrastructure permettant les circulations terrestres hors circulations ferroviaires, appartenant soit au domaine public ou privé d'une collectivité territoriale, soit à un individu ou groupement d'individus. Elle est composée d'un élément central, une chaussée, et des dépendances et accessoires permettant son maintien, la sécurité de ses usagers, et son agencement.* » (Pavard 2020).

*Auteur : Pavard 2021*

**Figure 2 – Exemples de conception d'un lexique sur la voirie**

**Comment décrire spatialement la voirie ?**

Au-delà, le manque d'une vision cohérente de la voirie empêche d'identifier son emprise précise et conduit à des estimations différentes. Ainsi, la surface de voirie de Paris est estimée entre 28km² (Ville de Paris 2020, Breteau 2016, Merlin 1985) et 15km² (CERTU 2013) selon que les dépendances sont inclues ou non (**Figure 3**).

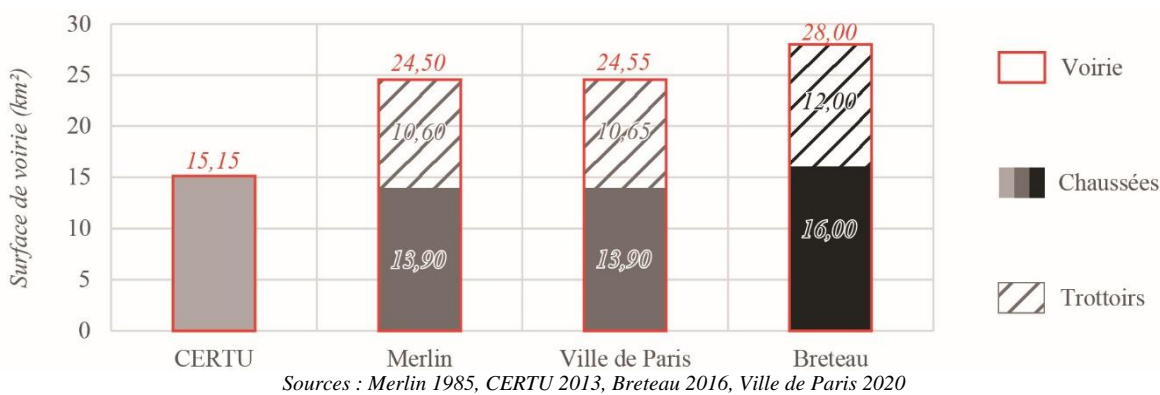

*Sources : Merlin 1985, CERTU 2013, Breteau 2016, Ville de Paris 2020*

**Figure 3 – Surfaces des emprises de voirie de Paris selon les définitions**

Les entreprises de travaux publics facturent leurs interventions selon la surface à couvrir. En effet, de celle-ci, découle la quantité de matériaux à utiliser. Par exemple, la couche visible d'une voirie, appelée revêtement, coûte entre 6 à 50 euros le mètre carré, tandis que la partie non visible coûte 240 à 520 euros le mètre carré (Cerema Ouest 2017). Un écart de 13 km² correspond à une variation d'au moins 3.1 milliards d'euros pour une opération de construction ou de rénovation.

Ensuite, l'information d'emprise doit être représentée et exploitable à travers des outils accessibles à tous les gestionnaires de voirie pour qu'ils organisent dans le temps et dans l'espace les opérations d'entretien. À ce titre, l'Observatoire National des Routes (ONR) a identifié que si l'État et les départements disposent des compétences, et des outils leur permettant de produire une connaissance fine de leur patrimoine, une majorité de communes rencontre des difficultés (IDRRIM 2017). **À partir de ces constats, deux questions émergent : quels outils utilisés pour la représentation spatiale de la voirie et comment en estimer les emprises ?**

D'abord, du côté des outils, les communes investissent depuis plusieurs décennies dans des logiciels informatiques de représentation spatiale appelés les Systèmes d'Information Géographiques (SIG). Aujourd'hui, ceux-ci sont largement déployés[1]. Il est alors plus réaliste pour des questions d'acceptabilité et de coût de s'appuyer sur les SIG plutôt que sur des outils à forts investissements. Ensuite, pour les méthodes de collecte, de construction et de gestion des informations liées à la voirie, il convient selon notre analyse d'utiliser des méthodes peu chronophages et s'appuyant au mieux sur des informations déjà construites. Plusieurs organismes se sont intéressés à des solutions de ce type et deux approchent en ont découlé :

1) le CERTU (2013) s'est focalisé sur l'usage de données utilisées habituellement pour les calculs d'itinéraires représentant les chaussées par des lignes suivant leur axe central pour estimer une largeur de voirie (**Figure 4 – a**) ;
2) l'Institut Paris Region s'est intéressé récemment, comme nous, à l'usage des informations cadastrales pour identifier l'espace public et par relation, les emprises de voirie (Delaville et al. 2021) (**Figure 4 – b**).

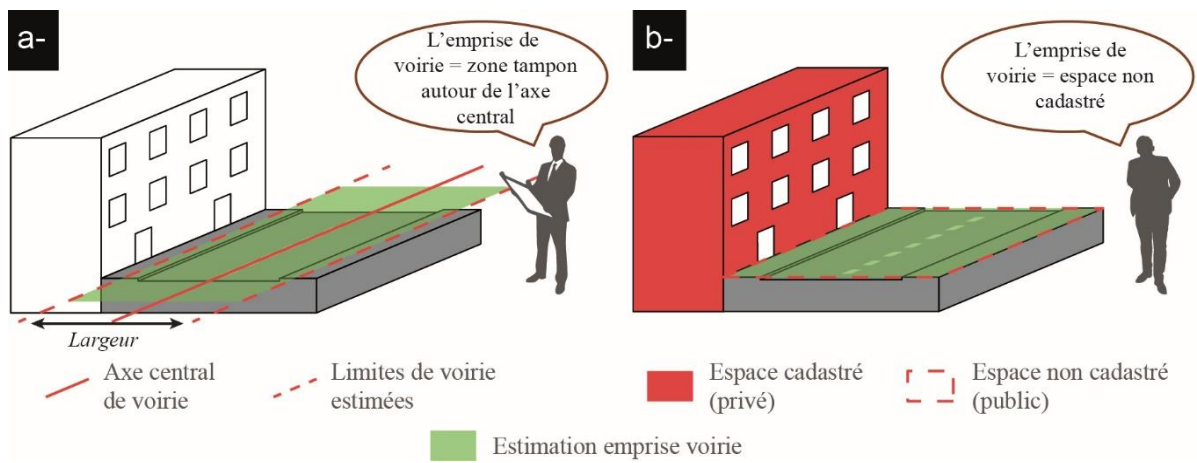

*Auteur : Pavard 2021*

**Figure 4 – Deux approches de construction des emprises de voirie**

Nous avons proposé une comparaison de ces deux approches en l'appliquant sur des cas d'études du territoire français (Pavard et al. 2021 ; Pavard et al. 2018) :

– le terrain de référence de nos travaux de recherche, la commune de Cachan au sud de Paris dans le Val-de-Marne (94) ;
– des terrains présentant des niveaux variés d'urbanisation : les communes de Bordeaux en Aquitaine, et de Paris ainsi que toutes les communes de l'Île de France.

Nos résultats ont permis d'aboutir à plusieurs constats :

---

[1] Déploiement des SIG dans les collectivités locales selon leur taille : 100% des communes d'au moins 100 000 habitants, 85% des communes d'au moins 50 000 habitants, 60% des communes d'au moins 10 000 habitants, 50% des communes d'au moins 5 000 habitants (IETI Consultants 2010 ; Mineau 2003).

1) La 1$^{ère}$ approche estime l'emprise des chaussées et non l'emprise de la voirie. Nous supposons que cela est lié à l'hétérogénéité des aménagements des voiries en ville.
2) La 2$^{ème}$ approche évalue bien l'emprise de la voirie. Des défauts sont liés à des aménagements spécifiques tels que les stationnements, ou les espaces verts en bordure.
3) La combinaison des deux approches permettrait de distinguer les chaussées des dépendances.
4) Les deux approches sont améliorables, en maitrisant mieux les règles d'attribution des largeurs de chaussées pour la 1$^{ère}$ approche, et en complétant par des levés terrains ou de l'analyse d'images aériennes la 2$^{ème}$ approche.

Finalement, les deux approches améliorées et combinées permettraient une description plus fine des éléments de voirie à l'aide d'informations structurelles utiles à leur bonne gestion.

**Comment décrire techniquement la voirie ?**

Les mobilités, les commerces, les loisirs, ou les liens sociaux, sont des fonctionnalités supportées par la voirie qui nécessitent son partage en sursol (Cerema 2018). Celui-ci est réalisé par une partition de l'espace à l'aide d'équipements, de marquages au sol, ou encore par une variation (nature, aspect et couleur) des matériaux de surface. Au-delà, la nature de l'activité supportée impacte les choix en terme de dimensionnement des éléments de la voirie en sous-sol. En effet, leur structure est adoptée pour répondre aux sollicitations induites par ces activités et pour limiter les impacts de la voirie sur ces activités. À titre d'exemples :

1) une chaussée supportant un trafic de véhicules nécessite une plus forte résistance mécanique qu'un trottoir ;
2) une canalisation métallique en sous-sol nécessite de choisir des matériaux non corrosifs pour la structure de l'élément de voirie qui l'accueille.

Ainsi, afin d'assurer l'entretien des voiries et de les aménager en cohérence avec les activités qu'elles supportent, les gestionnaires ont besoin de connaitre ces structures. Ce besoin est aussi juridique, puisque la règlementation en matière d'interventions sur réseaux enfouis sous voirie impose le remblayage d'une tranchée à l'identique. En d'autres termes, la nature et les épaisseurs des matériaux présents à l'origine doivent être respectés. Le caractère « identique » du remblai présuppose le besoin d'une connaissance préalable de l'infrastructure.

Enfin, la voirie évolue dans le temps et se dégrade sous l'effet de sollicitations :

1) mécaniques, liées aux trafics ;
2) thermiques et hydriques, liées aux intempéries ;
3) humaines liées à des interventions extérieures, comme les opérations implantant des nouveaux réseaux (fibre ou géothermie) et leur maintenance ;

Un manque d'entretien, notamment en surface, peut conduire à une diminution de la rugosité du revêtement et donc de la sécurité des usagers, ou à des défauts d'étanchéité. À terme, les infiltrations d'eau provoquent des désordres structurels irréversibles. Afin d'éviter une dégradation totale de l'infrastructure, les gestionnaires réalisent des opérations de maintenance sur la voirie.

En conclusion l'entretien de la voirie nécessite de connaitre l'état d'origine de la voirie (sa structure initiale), et de suivre son évolution (ses dégradations, ses maintenances). Il est donc indispensable d'associer une dimension temporelle aux informations structurelles venant enrichir la représentation spatiale de la voirie. **De là, émerge la question suivant : comment enrichir le référentiel spatial de voirie par des informations techniques ?**

Lors de nos recherches, nous avons constaté que les communes disposent peu, ou dans un format non adapté, des informations essentielles (structures et dégradation). Ce manque est lié à des difficultés de collecte, de référencement ou encore à des pertes de documents, et a plusieurs impacts (Pavard et al. 2019) :

1) Un manque d'informations sur les structures nécessite de réaliser des fouilles (carottage), lesquelles entrainent des dépenses. Ensuite, une information manquante ou mal référencée peut induire des incidents lors d'interventions à proximité de réseaux (La Dépêche 2008, Le Figaro 2008 ; Le Monde 2007 ; Libération 2007).
2) Un manque d'informations sur l'évolution des voiries entraine une planification des maintenances non adaptée. Cela a pour effet de mal allouer les financements et de ne pas intervenir à temps sur des dégradations.

**De ces éléments découlent trois autres questions : quelles sont les informations techniques essentielles, comment les collecter et les associer au référentiel ?** Afin d'aider les gestionnaires, et notamment les moins avancés, pour des raisons de manque de moyens techniques, financiers ou humains, nous avons proposé :

1) une hiérarchisation et une organisation des informations afin de les lier au référentiel spatial conçu ;
2) des méthodes simples de collecte d'informations telles qu'une levée visuelle pour les revêtements et les dégradations par pas de 10 mètres ;
3) des collectes d'informations géolocalisées sur les structures de voirie lors d'opérations d'excavation pour maintenance de réseaux par exemple.

Finalement, nous avons appliqué l'ensemble du processus de construction du référentiel spatial de voirie et son enrichissement sur le terrain de référence de Cachan afin d'en démontrer ces apports.

**En savoir plus**

Pavard, Antonin. 2020. *Optimiser la gestion conjointe de la voirie et des réseaux enterrés à l'aide de la géomatique : conception d'un référentiel spatial de voirie : Constructibilité à l'interface entre route et réseaux enterrés*, thèse de sciences de l'ingénieur, Université Paris-Est Marne-la-Vallée. Disponible à URL suivant : https://tel.archives-ouvertes.fr/tel-03270655/.

**Antonin Pavard** est diplômé de l'Université Paris-Est Marne-la-Vallée, docteur en sciences de l'ingénieur mention géomatique, et enseignant-chercheur à l'Institut de Recherche en Constructibilité (IRC) de l'Ecole Spéciale des Travaux Publics (ESTP-Paris). Il a soutenu la thèse « Optimiser la gestion conjointe de la voirie et des réseaux enterrés à l'aide de la géomatique : conception d'un référentiel spatial de voirie ». Ses thèmes de recherche s'articulent autour de la modélisation spatio-temporelle des territoires pour répondre à des enjeux de mobilité, de gestion des infrastructures urbaines, ou encore de gestion des risques naturels et technologiques.

**Anne Dony** est ingénieur de formation et docteur en sciences physiques par une thèse dans le domaine des matériaux bitumineux. Après 17 ans au sein du service technique d'une entreprise routière, elle travaille depuis 2004 en tant qu'enseignant-chercheur au sein de l'ESTP et est responsable de la thématique matériaux de chaussées au sein de l'IRC, Institut de Recherche en constructibilité. Ses thématiques de recherche s'articulent autour des matériaux de construction routière, les enjeux liés aux transitions climatiques et environnementales (économie circulaire, abaissement des températures…) et une prise en compte des nouvelles fonctionnalités de la route.

**Patricia Bordin** est docteur en sciences de l'information géographique. Elle développe des recherches en traitements et modélisations géographiques avec différents laboratoires ou dans le cadre de GéoSpective la structure qu'elle a créée en 2010. Avec le temps et de nombreuses collaborations avec différents gestionnaires de réseaux, elle a acquis une expertise particulière dans le domaine de leurs applications aux réseaux.